\documentclass[referee]{raa}            

\usepackage{graphicx,times}             
\usepackage{natbib}

\usepackage[papersize={8.5in,11in}]{geometry}

\providecommand{\bm}[1]{\mbox{\boldmath$#1$\unboldmath}}

\begin{document}

\title{Preparation for CSST: Star-Galaxy Classification using a Rotationally-invariant Supervised Machine-learning Method}
\volnopage{Vol.0 (2024) No.0, 000--000}
\author{Shiliang Zhang \inst{1}
\and Guanwen Fang \inst{1}
\and Jie Song \inst{2,3}
\and Ran Li \inst{4}
\and Yizhou Gu \inst{5}
\and Zesen Lin \inst{6}
\and Chichun Zhou \inst{7}
\and Yao Dai \inst{1}
\and Xu Kong \inst{2,3}
 }

\institute{Institute of Astronomy and Astrophysics, Anqing Normal University, Anqing, 246133, China; {\it wen@mail.ustc.edu.cn};\\
\and Deep Space Exploration Laboratory/Department of Astronomy, University of Science and Technology of China, Hefei, 230026, China; {\it xkong@ustc.edu.cn};\\
\and School of Astronomy and Space Science, University of Science and Technology of China, Hefei, 230026, China;\\
\and National Astronomical Observatories, Chinese Academy of Sciences, Beijing, 100101, PR China;\\
\and Tsung-Dao Lee Institute and Key Laboratory for Particle Physics, Astrophysics and Cosmology, Ministry of Education, Shanghai Jiao Tong University, Shanghai, 200240, China;\\
\and Department of Physics, The Chinese University of Hong Kong, Shatin, N.T., Hong Kong, S.A.R., China;\\
\and School of Engineering, Dali University, Dali, 671003, China; \\
}

\abstract{Most existing star-galaxy classifiers depend on the reduced information from catalogs, necessitating careful data processing and feature extraction. In this study, we employ a supervised machine learning method (GoogLeNet) to automatically classify stars and galaxies in the COSMOS field. Unlike traditional machine learning methods, we introduce several preprocessing techniques, including noise reduction and the unwrapping of denoised images in polar coordinates, applied to our carefully selected samples of stars and galaxies. By dividing the selected samples into training and validation sets in an 8:2 ratio, we evaluate the performance of the GoogLeNet model in distinguishing between stars and galaxies. The results indicate that the GoogLeNet model is highly effective, achieving accuracies of 99.6\% and 99.9\% for stars and galaxies, respectively. Furthermore, by comparing the results with and without preprocessing, we find that preprocessing can significantly improve classification accuracy (by approximately 2.0\% to 6.0\%) when the images are rotated. In preparation for the future launch of the China Space Station Telescope (CSST), we also evaluate the performance of the GoogLeNet model on the CSST simulation data. These results demonstrate a high level of accuracy (approximately 99.8\%), indicating that this model can be effectively utilized for future observations with the CSST.
 }

\authorrunning{Shiliang Zhang et al. } 
\titlerunning{Preparation for CSST } 
\maketitle

\section{Introduction} \label{sect:intro}

In astronomy, precise differentiation between stars and galaxies is paramount due to their representation of distinct astrophysical phenomena. For instance, the systematic contribution resulting from the cross-contamination of star and galaxy samples could significantly impact the field of ``precision cosmology'' (e.g., \citealt{rossAmelioratingSystematicUncertainties2011a, thomasExcessClusteringLarge2011, soumagnacStarGalaxySeparation2015, 10.1093/mnras/sty2579}). This issue will become increasingly critical in future astronomical research, as the upcoming large-field sky surveys, such as those conducted by the Chinese Space Station Telescope (CSST) \citep{zhanConsiderationLargescaleMulticolor2011, zhan2018overview}, the Euclid Space Telescope \citep{euclidcollaborationEuclidPreparationEuclid2022}, and the Roman Space Telescope \citep{spergelWideFieldInfrarRedSurvey2015}, will yield imaging of millions to billions of stars and galaxies. This necessitates the development of methods to accurately and rapidly distinguish between stars and galaxies.

Several methods are currently available to address this issue. The first classification method is morphology-based, involving the determination of an optimal threshold in the space of observable image properties (e.g., \citealt{10.1093/mnras/176.2.265, kronPhotometryCompleteSample1980, Leauthaud_2007, skelton3DHSTWFC3SELECTEDPHOTOMETRIC2014, 10.1093/mnras/sty2579, refId0}). This method is predicated on the assumption that stars are point-like sources, while galaxies are extended sources. Consequently, stars and galaxies can be differentiated by their distribution in a size-magnitude diagram (e.g., \citealt{Leauthaud_2007, skelton3DHSTWFC3SELECTEDPHOTOMETRIC2014}). Another approach is color-based, leveraging the distinct spectral shapes of stars and galaxies. This method allows for the differentiation between stars, galaxies, and quasars through color-color diagrams (e.g., \citealt{baldryGalaxyMassAssembly2010, sagliaPHOTOMETRICCLASSIFICATIONSERVER2012}). Combining these two methods is also feasible to maximize the utilization of available data  (e.g., \citealt{10.1093/mnras/stv1608, soumagnacStarGalaxySeparation2015, 10.1093/mnras/stw2672}).

However, these empirical methods rely on reduced summary information, which could be challenging to obtain. Transforming astronomical images into suitable features necessitates careful engineering and considerable domain expertise. Machine learning, particularly Convolutional Neural Networks (CNNs), offers a potent alternative by facilitating the direct extraction of image features (e.g., \citealt{10.1093/mnras/stv632, Huertas-Company_2015,10.1093/mnras/sty3232}), circumventing the need for manual feature design. This approach enables a more efficient and accurate classification between stars and galaxies.

The application of machine learning techniques in classifying stars and galaxies was pioneered by \cite{1992AJ....103..318O}, and it has since become an integral component of widely used software packages such as \texttt{SExtractor} \citep{1996A&AS..117..393B}. Subsequently, various successful implementations have emerged to tackle this problem, including decision trees, support vector machines, and classifier ensemble strategies (e.g., \citealt{1995AJ....109.2401W, Suchkov_2005, Ball_2006, Vasconcellos_2011, SEVILLANOARBE201564, Fadely_2012, 2017MNRAS.464.4463K2}). However, studies have demonstrated that traditional CNN-based machine learning algorithms often exhibit poor robustness to signal-to-noise ratios ($\rm S/N$) and image rotations since noise and rotations may break the image features (e.g., \citealt{nazare2018deep, liu2020networks, 7560644, Cabrera-Vives_2017, 8462057, 8445665}). Efforts have been made to address these challenges. For instance, noise reduction has been identified as an effective method for mitigating the impact of image $\rm S/N$ (e.g., \citealt{nazare2018deep, 2023AJ....165...35F}). Regarding the influence of image rotation, our previous studies have demonstrated that APCT (Adaptive Polar Coordinate Transformation) can effectively overcome the challenges posed by image rotation (e.g., \citealt{2023AJ....165...35F, Dai_2023, Song2024USmorphAU}). These approaches can significantly enhance the accuracy of machine learning models in recognizing image features.

In this study, we employ the GoogLeNet algorithm \citep{7298594} to the Hubble Space Telescope (HST) I-band images of the Cosmic Evolution Survey (COSMOS) field to classify stars and galaxies.
The GoogLeNet algorithm, previously validated for high-resolution images (e.g., \citealt{2023AJ....165...35F, Dai_2023}), is found to effectively distinguish between stars and galaxies after preprocessing the images with noise reduction and APCT. Comparing the results obtained with and without preprocessing, we observe a significant improvement in classification accuracy. Furthermore, in anticipation of the future launch of the CSST, we evaluate the robustness of our framework using the simulated CSST data. The results demonstrate an accuracy approaching 99.8\%, indicating the suitability of this framework for future observations with the CSST.

The structure of this paper is as follows. In Section \ref{sec:data}, we outline how we select our samples of stars and galaxies. In Section \ref{sec:method}, we describe our data preprocessing method and the GoogLeNet algorithm. The classification results are presented in Section \ref{sec:results}. In Section \ref{sec:CSST}, we evaluated the performance of our framework on the CSST simulated data. Finally, conclusions are provided in Section \ref{sec:summary}.

\section{Data Set and Sample Selection} \label{sec:data}
\subsection{COSMOS Field} \label{subsec:COSMOS}

The COSMOS field \citep{2007ApJS..172....1S} has significantly advanced our understanding of the Universe by providing deep data covering a wide wavelength range from radio to X-rays. In this study, we utilize the high-resolution imaging data acquired from HST with the Advanced Camera for Surveys (ACS) in the F814W band. This dataset comprises approximately 590 pointings and encompasses an area of 1.64 square degrees within the COSMOS field, making it the largest contiguous survey conducted by the HST/ACS to date. Its extensive coverage ensures a sufficient number of stars, rendering it suitable for our investigation. The original images were processed by \cite{koekemoerCOSMOSSurveyHubble2007} using the \texttt{MultiDrizzle} package \citep{2003hstc.conf..337K}. The final mosaic images have a pixel scale of $0\farcs03$, and the $5\sigma$ depth is 27.2 AB magnitude for point source observations within an aperture diameter of $0\farcs24$. Subsequent analyses in this study are conducted based on these high-resolution mosaic images.

\subsection{COSMOS2020 Catalog} \label{subsec:catalog}
The dataset utilized in this study is derived from the COSMOS2020 catalog \citep{2022ApJS..258...11W}, which offers reliable photometric estimations for approximately 1.7 million objects spanning from far ultraviolet to near-infrared wavelengths. These objects are detected from the ``chi-square'' $iz Y J H K_s$ image. The luminosity is estimated using two independent extraction methods: (1) the Classic catalog employing \texttt{SExtractor}, and (2) the Farmer catalog employing parametric modeling via the \texttt{Tractor} package \citep{2016ascl.soft04008L}. Additionally, photometric redshifts and other physical parameters are also estimated through spectral energy distribution (SED) fitting using two different approaches, namely \texttt{LePhare} \citep{ilbertAccuratePhotometricRedshifts2006} and \texttt{EAZY} \citep{brammerEAZYFastPublic2008}, for both catalogs. \cite{2022ApJS..258...11W} provided a detailed comparison among these different estimation methods, demonstrating their high consistency. We opt for the Classic catalog with redshifts estimated using \texttt{LePhare} in this study.

Additionally,  \cite{2022ApJS..258...11W} distinguished stars from galaxies by combining morphological and SED criteria. Point-like objects form a tight sequence in the half-light radius versus magnitude space, thus, \cite{2022ApJS..258...11W} classified all bright sources on this sequence as stars using morphological information obtained from HST/ACS ($I$ band) and Subaru/HSC ($i$ band) images. They also compared the best-fit $\chi^2$ values obtained using stellar ($\chi^2_{star}$) and galaxy ($\chi^2_{gal}$) templates during SED fitting with \texttt{LePhare}. Objects with $\chi^2_{star}$ smaller than $\chi^2_{gal}$ were classified as stars. In the COSMOS2020 catalog, the parameter ``$\rm lp_{type}$'' indicates the final classification results between stars and galaxies, with galaxies assigned $\rm lp_{type}$ = 0 and stars assigned $\rm lp_{type}$ = 1.

\subsection{Selection of Galaxies for Analysis} \label{subsec:galaxy selection} 

\begin{figure}[t]
\centering
\includegraphics[width=0.6\columnwidth]{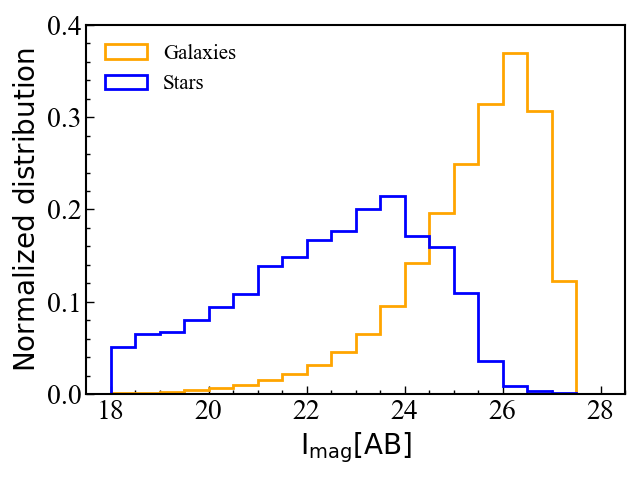}
\caption{Distributions of I-band magnitudes for stars (blue) and galaxies (yellow), separated by the $\rm lp_{type}$ parameter, in the COSMOS field. In this work, we only consider objects brighter than 25 mag.\label{fig:fig01}}
\end{figure}

In this study, we obtain the photometric information from the Classic-version catalog, along with various other physical parameters estimated using \texttt{LePhare}. Our galaxy samples are selected based on the following criteria: (1) $\rm lp_{type}$ = 0, indicating the object is classified as a galaxy; (2) $I_{\rm mag} < 25$, excluding objects that are too faint; (3) $0.2 < z < 1.2$, ensuring that the galaxies are observed in the rest-frame optical wavelength; (4) $\rm FLAG_{COMBINE}$ = 0, indicating that the photometric measurements are not contaminated by bright stars. Additionally, we exclude all sources with bad pixels. Finally, we randomly select 60,000 galaxies from the remaining samples as our galaxy sample. In Figure \ref{fig:fig01}, we present the distribution of I-band magnitudes for stars and galaxies (as separated by the $\rm lp_{type}$ parameter) in the COSMOS field. 

\subsection{Selection of Stars for Analysis} \label{subsec:stellar selection}

To accurately evaluate the performance of the GoogLeNet model in distinguishing between stars and galaxies, it is essential to obtain clean samples of stars. This is particularly important because the number of stars in this field is relatively small compared to the number of galaxies, and contamination could significantly affect the accuracy of star classification. Therefore, in addition to restricting the selection criteria to mirror those used for galaxy identification, which include: (1) $\rm lp_{type}$ = 1, (2) $I_{\rm mag} < 25$, and (3) $\rm FLAG_{COMBINE}$ = 0, we also take further steps to ensure the reliability of our star sample.

\cite{2022ApJS..258...11W} had already distinguished stars from galaxies using the half-light radius and magnitude for bright sources. Any bright objects ($I < 23$ mag for HST images and $i < 21.5$ mag for HSC images) falling on the point-like source sequence were classified as stars by . \cite{2022ApJS..258...11W}. Considering that our sample includes some fainter sources, we also employed similar methods to obtain a purer sample of stars with I-band magnitudes $I_{\rm mag} < 25$.

\begin{figure}
	\centering
	\includegraphics[width=0.6\columnwidth]{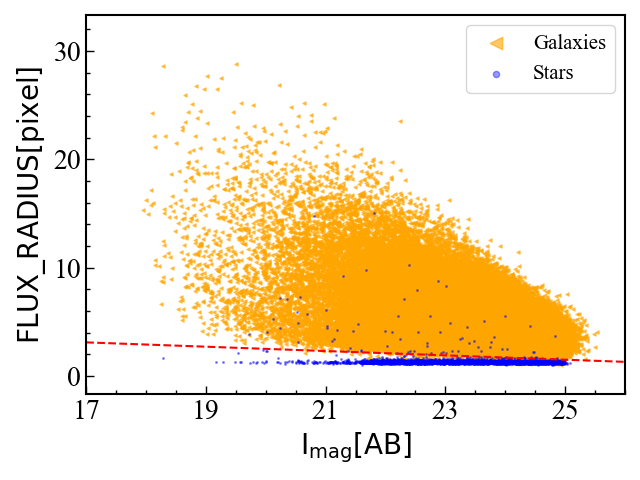}
	\caption{Distributions of FLUX\_RADIUS and $I_{\rm mag}$ for our samples, separated by $\rm lp_{type}$. Galaxies and stars are represented by yellow triangles and blue points, respectively. It is evident from this figure that stars form a tight sequence. We separate the stars and galaxies using a custom criterion, shown as a red line in this figure. All sources above this line are removed from our star sample.\label{fig:fig02}}
\end{figure}

Firstly, \cite{2016ApJS..224...24L} demonstrated that stars form a tight sequence ($z^{++} - K_{\rm s} < (B - z^{++}) \times 0.3 - 0.2$) in the $BzK_{\rm s}$ color-color diagram. Following their criteria, we consider only stars that are not too far from this sequence. Additionally, we use the size-magnitude diagram to examine the contamination of the star sample we selected earlier. The results are illustrated in Figure \ref{fig:fig02}, where the distributions of galaxies and stars are represented by yellow triangles and blue points, respectively. Here, we do not attempt to obtain the half-light radius of each object by fitting their light distribution with different models. Instead, we directly use the FLUX\_RADIUS provided by \texttt{SExtractor}. Some samples classified as stars by $\rm lp_{type} = 1$ are located in the region of extended sources in this figure, even for objects brighter than 23 magnitudes in I band. Considering the different definitions of the size used by us and \cite{2022ApJS..258...11W}, the presence of these bright extended sources with $\rm lp_{type} = 1$ is reasonable. We eliminate potential extended sources from our stellar sample with a custom criterion, $\rm FLUX\_RADIUS > -0.2 \times I_{\rm mag} + 6.5$, shown as a red dashed line in Figure \ref{fig:fig02}. Any sources located above this line are excluded from the following analysis. A total of 91 stars have been excluded. Moreover, the Classic catalog also provides the class\_star parameter, which is estimated by \texttt{SExtractor} to distinguish stars from galaxies. Any ambiguous star samples showing class\_star values lower than 0.85 are also removed. We have summarized the process of selecting galaxy and star samples in Table \ref{tbl-1}.

To verify the purity of our selected star sample, we present the distribution of the selected stars and galaxies in the $gzK_{\rm s}$ color-color diagram in Figure \ref{fig:fig03}. Numerous studies have shown that stars follow a tight sequence in this color-color space (e.g., \citealt{arcila2013numbers, 2022ApJS..258...11W}). As shown in Figure \ref{fig:fig03}, our selected star samples also tightly follow this sequence, demonstrating the reliability of our selection method. Utilizing the aforementioned methods, we ultimately obtain a clean sample comprising 7,102 stars and 60,000 galaxies.

\begin{table*}[htbp]
    \centering
    \caption{Selection of galaxy and star samples}\label{tbl-1}
    \begin{tabular}{ccc}
    \hline\hline\noalign{\smallskip}
    Step & Galaxy Selection & Star Selection\\   
    \noalign{\smallskip}\hline\noalign{\smallskip}
    1 & \begin{tabular}{c} $I_{\rm mag} < 25$ \\ $\rm FLAG_{COMBINE}$ = 0 \end{tabular} & \begin{tabular}{c} $I_{\rm mag} < 25$ \\ $\rm FLAG_{COMBINE}$ = 0 \end{tabular} \\
    2 & $\rm lp_{type} = 0$ & $\rm lp_{type} = 1$ \\
    3 & $0.2 < z < 1.2$ & $z^{++} - K_{\rm s} < (B - z^{++}) \times 0.3 - 0.2$ \\
    4 & -- & $\rm FLUX\_RADIUS < -0.2 \times I_{\rm mag} + 6.5$ \\
    5 & -- & class\_star $< 0.85$ \\
    \noalign{\smallskip}\hline
    \end{tabular}
\end{table*}

\section{Preprocessing of Data and SML Models} \label{sec:method}

\subsection{Noise Reduction}\label{subsec:noise reduction}

Several previous studies have shown that noise can disrupt the features extracted from images by machine learning, potentially leading to incorrect classification results \citep{2023AJ....165...35F}. Noise reduction has been identified as a viable solution to this issue. \cite{masci2011stacked}
demonstrated that image quality can be significantly enhanced by extracting the primary features of images and subsequently reconstructing the images from these extracted features. In this work, we adopt Convolutional Autoencoders (CAE) for noise reduction, which has been proven effective in many other studies (e.g., \citealt{2022AJ....163...86Z, 2023AJ....165...35F, Dai_2023, Song2024USmorphAU}). The specific approach we use follows that of \cite{Song2024USmorphAU}.

First, we crop our samples into cutouts of $100 \times 100$ pixels, centering the objects within the image. We then applied CAE to reduce the noise in the images. The CAE configurations we adopt are identical to those used by \cite{Song2024USmorphAU}. For more details, we refer readers to that paper. In Figure \ref{fig:fig04}, we present the results of image denoising for some randomly selected samples. The first column showcases the raw images of four randomly selected objects, while the second column displays the corresponding denoised counterparts. It is evident that the CAE significantly enhances image quality without breaking the main features.

\subsection{Adaptive Polar Coordinate Transformation} \label{subsec: APCT}

\begin{figure}
	\centering
	\includegraphics[width=0.6\columnwidth]{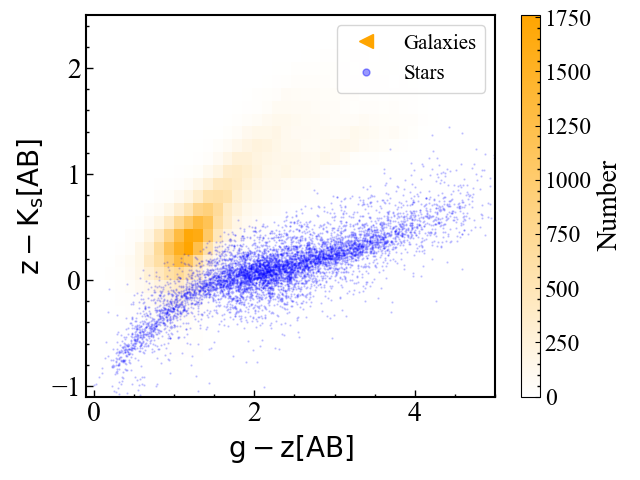}
	\caption{Distributions of our final selected galaxy and star samples in the $\rm gzK_s$ diagram. Galaxies are represented by the yellow 2D histogram with color representing the number in each bin, while stars are represented by blue points. The stars are clearly located on a distinct sequence, indicating that we have obtained clean star samples.\label{fig:fig03}}
\end{figure}

In the field of astronomy, the classification of stars and galaxies should be independent of image rotations. However, existing supervised machine learning (SML) algorithms, particularly those based on CNNs, may misclassify them when images are rotated. Various methods have been proposed to mitigate this issue, including data augmentation techniques and traditional polar coordinate transformation (e.g., \citealt{8462057, liu2019automated, Mo2022RICCNNRC}). However, data augmentation is very computationally intensive and inefficient, making it challenging to apply to future large-scale sky surveys. Polar-coordinate transformation is more efficient, but conventional polar-coordinate transformation cannot perfectly convert the rotations of the raw images into new images due to the integer coordinates of the pixels (as shown in Figure \ref{fig:fig03} of \cite{2023AJ....165...35F}). To overcome this shortcoming, we have proposed the APCT method in our previous works (e.g., \citealt{2023AJ....165...35F, Dai_2023, Song2024USmorphAU}).

\begin{figure}[t]
	\centering
	\includegraphics[width=0.9\linewidth]{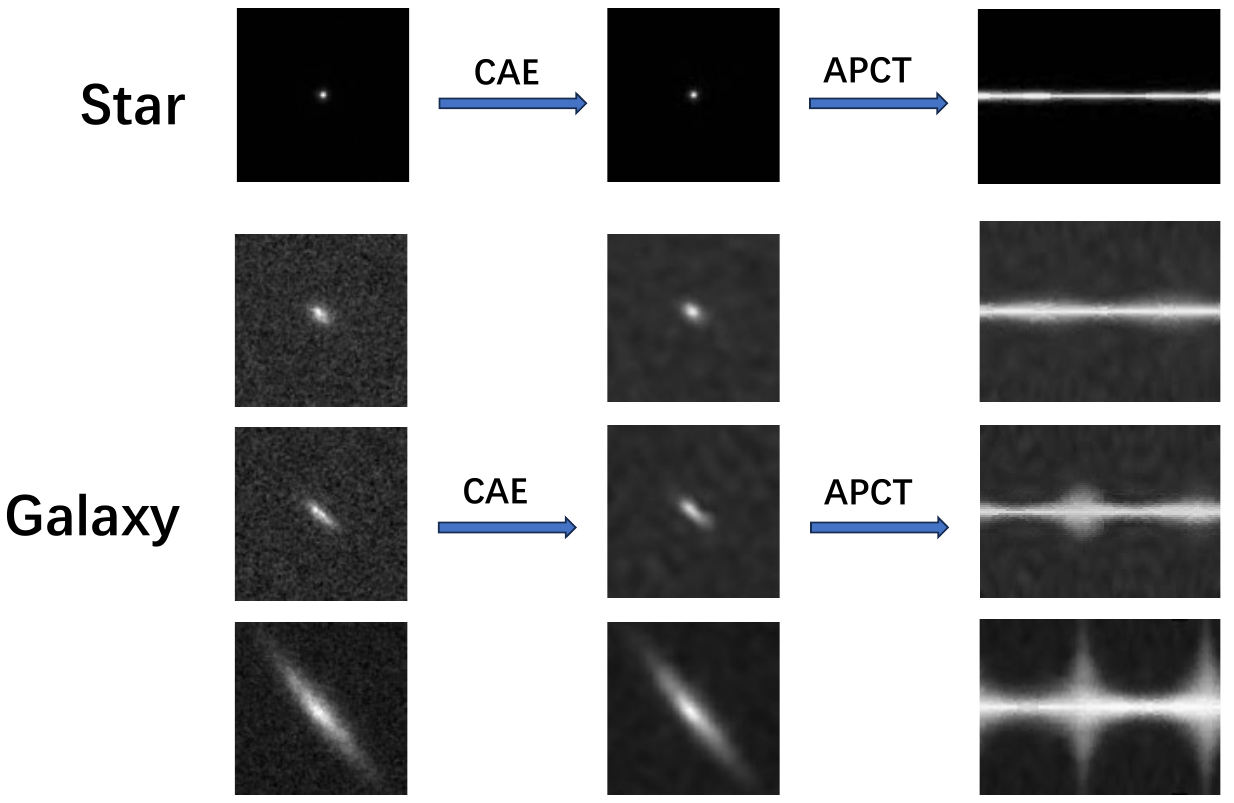}
	\caption{Examples of our preprocessing method. The first column shows images of four randomly selected samples, the second column displays the corresponding denoised images, and the third column presents the results after transformation into polar coordinates.\label{fig:fig04}}
\end{figure}

In brief, the APCT method employs rotation-invariant polar axes, making it more robust to image rotation. This involves selecting the pixels with the highest and lowest flux values as the brightest and darkest points. The line connecting the brightest to the darkest point is designated as the polar axis for the polar-coordinate system. This designation ensures the polar axis remains unaffected by rotation. Next, we rotate the axis counterclockwise in increments of 0.05 radians. For each discrete rotation, the axis passes through many pixels of the original image. By stacking pixels along this rotating axis during rotation, a new image is obtained. Considering that CNNs are more sensitive to information in the center of images, we apply a mirroring process to the transformed images. The third column in Figure \ref{fig:fig04} shows the corresponding results after APCT for our randomly selected samples.

\subsection{GoogLeNet Algorithm} \label{subsec:GoogLeNet}

In 2014, GoogLeNet \citep{7298594} achieved remarkable results in the ImageNet image classification challenge, demonstrating its effectiveness in image classification tasks. The GoogLeNet architecture features nine sequentially stacked inception modules. Each inception module employs parallel applications of convolutions with kernel sizes of $1\times 1$, $3\times 3$, and $5\times 5$. This configuration allows the network to cover a broader area of the input images while preserving fine detail in smaller regions. By expanding both the depth and width of the network, it achieves enhanced parameter efficiency while concurrently minimizing the total number of parameters. This advancement significantly contributes to the model's performance and computational efficiency. In our previous work,  \cite{2023AJ....165...35F} tested the performance of three different machine learning frameworks and found that the GoogLeNet model performs best for these deep high-resolution images. Therefore, we also adopt the GoogLeNet model in this work. The algorithmic flow of the GoogLeNet used in our study is illustrated in Figure \ref{fig:fig05}. The model parameters used in this work are the same as those in  \cite{2023AJ....165...35F}.

\section{Results and Analysis}\label{sec:results}
\subsection{Classification Results of COSMOS Data}\label{subsec:classification result}

In Section \ref{sec:data}, through meticulous selection, we have obtained reliable samples of stars and galaxies. Using the preprocessed images of these samples, we can evaluate the performance of the GoogLeNet algorithm in distinguishing between stars and galaxies. To avoid overfitting, we randomly divide our sample into training (48,040 galaxies and 5,640 stars) and validation (11,960 galaxies and 1,462 stars) sets in an 8:2 ratio and estimate the precision and recall rates based on this validation set.

The results of our classification are presented in Figure \ref{fig:fig06}, where the left panel and right panel show the recall and precision rates, respectively. The recall rates for both stars and galaxies exceed 99.0\%, while the precision rates are higher than 99.5\%. Additionally, when considering the specific number of classification errors, only 6 galaxies and 12 stars are misclassified by the GoogLeNet model, as presented in Table \ref{tab:1}. These results demonstrate the GoogLeNet model's robust performance in classifying star and galaxy images, with a low probability of misclassification between them.

\begin{figure}[htp!]
	\centering
	\includegraphics[width=0.44\linewidth]{ms2024-0205fig5.png}
	\caption{The structure of GoogLeNet used in this work is outlined in this figure: there are 22 layers in total, with no fully connected layers. The operations performed by each layer are represented in the boxes.\label{fig:fig05}}
\end{figure}

\begin{figure}
	\centering
	\includegraphics[width=0.9\linewidth]{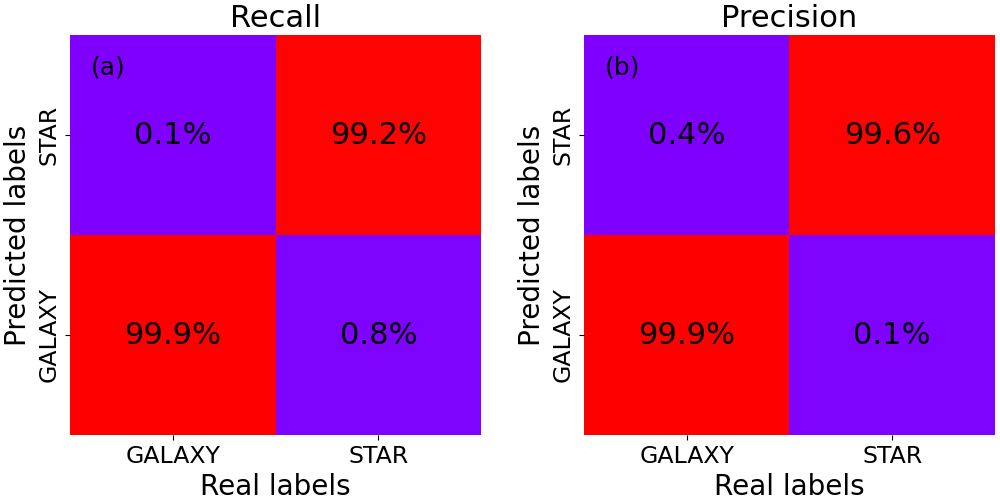}
	\caption{The recall (panel a) and precision (panel b) rates estimated from the validation set of the GoogLeNet model are depicted. The overall accuracy rates for both stars and galaxies exceed 99.0\%, indicating that our framework demonstrates excellent performance in classifying stars and galaxies.\label{fig:fig06}}
\end{figure}

Some previous works have also utilized machine learning for star-galaxy classification. For example, using data from the J-PLUS Early Data Release,  \cite{refId0} implemented a Bayesian classifier for morphological star-galaxy classification based on Probability Density Function analysis. They provided reliable probabilities for statistical analysis of 150,000 stars and 101,000 galaxies, achieving a completeness of approximately 95.0\% and a contamination of approximately 5.0\% for both stars and galaxies up to $\rm r\sim 20$. Compared to their results, our study demonstrates the superior performance of the GoogLeNet model, with significantly higher precision and recall rates and a minimal misclassification rate, emphasizing the model's robustness and reliability in classifying stars and galaxies.

\begin{table} 
\centering
\caption{The specific classification results of the validation set.}\label{tab:1} 
\setlength{\tabcolsep}{5mm} 
\begin{tabular}{cccc} 
\hline\hline\noalign{\smallskip}	
Pred/Real & Galaxies & Stars & Total \\ 
\noalign{\smallskip}\hline\noalign{\smallskip}
Stars & 6 & 1450 & 1456 \\ 
Galaxies & 11954 & 12 & 11965 \\
\noalign{\smallskip}\hline
\end{tabular} 
\end{table}

\subsection{The Importance of Preprocessing}

In Section \ref{sec:method}, we have thoroughly explained the necessity of preprocessing. The importance of noise reduction has been carefully discussed in many previous studies. Regarding APCT, 
 \citep{2023AJ....165...35F} demonstrated its significance in galaxy morphology classification. When the test set is rotated, the overall accuracy significantly decreases from over 95.0\% to approximately 83.0\% without APCT. However, after performing APCT, the largest difference is only 1.5\% no matter how many degrees the test set is rotated. In this section, we will further investigate the importance of our preprocessing method in the classification of stars and galaxies.

\subsubsection{T-SNE Test}

\begin{figure*}
	\centering
	\includegraphics[width=0.93\linewidth]{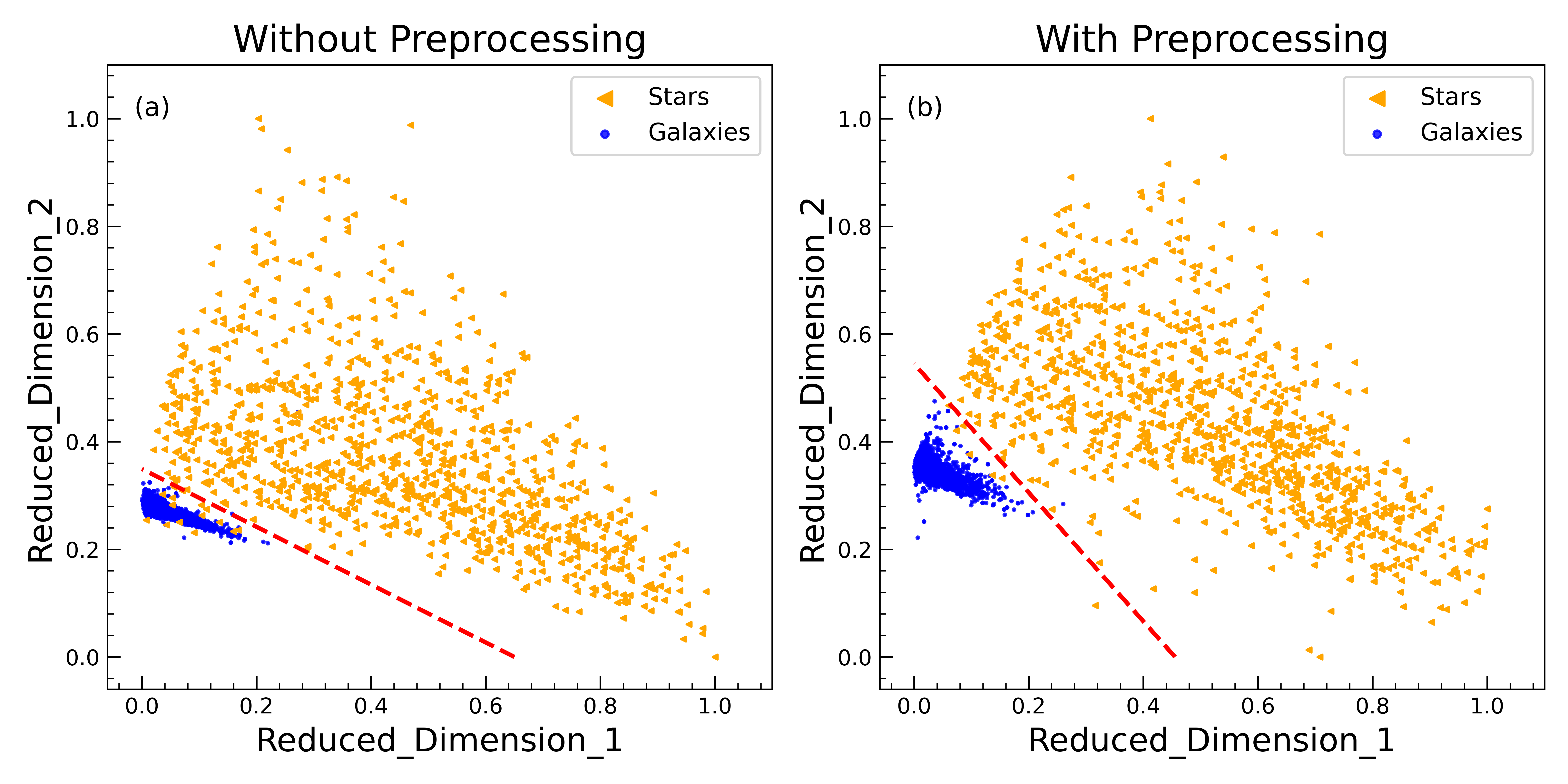}
	\caption{The t-SNE diagram depicts our randomly selected 5.0\% samples, with the left panel illustrating the distribution of stars and galaxies without preprocessing, and the right panel showing the results after preprocessing. Using the data after t-SNE dimensionality reduction, a SVM classification on this feature plot is applied and the result is shown as the red dashed line in the figure. It can be seen clearly that after preprocessing, the galaxies and stars are indeed more distinctly separated on this t-SNE plot.\label{fig:fig07}}
\end{figure*}

The t-Distributed Stochastic Neighbor Embedding (t-SNE) technique provides us with an effective method for mapping high-dimensional data into a lower-dimensional space, thereby transforming distributions into dimensions suitable for examination (e.g., \citealt{vandermaaten, van2014accelerating, wattenberg2016use}). In Figure \ref{fig:fig07}, we present the t-SNE diagram for our samples before and after preprocessing in panels (a) and (b), respectively. To facilitate visualization, we only present the results of a randomly selected 5.0\% subset of our entire sample. We have also examined the distribution of all samples and found that the results remained consistent.

It is generally believed that images with similar morphologies should have similar features, leading to distinct separable boundaries between different classes. So stars and galaxies should be far apart in this diagram. In both panels of Figure \ref{fig:fig07}, stars and galaxies exhibit clear separable boundaries, indicating that we have indeed provided viable prior samples for the GoogLeNet model. Although there is a slight overlap between stars and galaxies in the t-SNE diagram, this can be reasonable given the projection effects. Moreover, compared to panel (a), the boundaries between the two classes appear more distinct in panel (b), indicating the effectiveness of our preprocessing steps.


Since t-SNE can only qualitatively describe the effects of preprocessing, we have also applied an Support Vector Machines (SVM) classification to this feature plot to make a quantitative
demonstration. The boundary lines between galaxy and star samples are estimated with SVM technique, and the results shown as red dashed lines in Figure \ref{fig:fig07}. Since the lower left corner of each panel is predicted as star-like samples, and the upper right corner is predicted as galaxy-like samples by SVM classification, we calculate the accuracy and recall rates for galaxies and stars for this classification. Without preprocessing, the precision and recall rates for galaxies (stars) are 99.7\% and 98.6\% (98.6\% and 99.7\%), respectively. After preprocessing, the precision and recall rates improved to 99.9\% and 99.4\% (99.5\% and 99.9\%) for galaxies (stars). This indicates that galaxies and stars are indeed more distinctly separated in this t-SNE plot, indicating the effectiveness of our preprocessing process.


\subsubsection{The Effectiveness of Preprocessing}

\begin{figure*}
	\centering
	\includegraphics[width=0.93\linewidth]{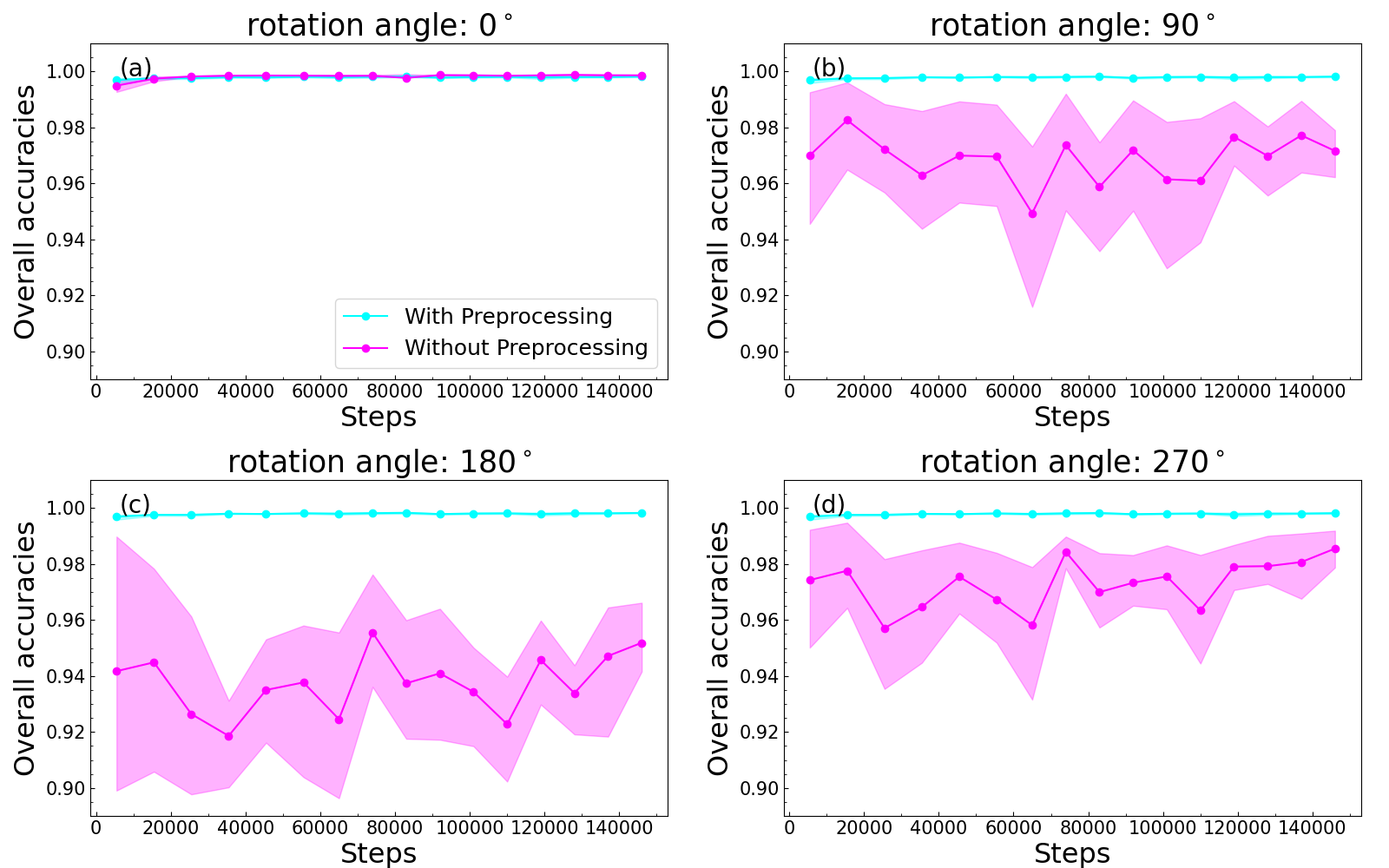}
	\caption{The accuracy as a function of training steps when images in the validation set are rotated with different angles is presented in four panels: panel (a) ($0^\circ$), panel (b) ($90^\circ$), panel (c) ($180^\circ$), and panel (d) ($270^\circ$). In each panel, the cyan and magenta lines represent the median accuracy rate in each step bin estimated with and without preprocessing, respectively. The shaded regions represent the corresponding 16th and 84th percentiles of accuracy rate at each step bin.\label{fig:fig08}}
\end{figure*}

To quantitatively demonstrate the effectiveness of preprocessing, we calculate the overall accuracy of the GoogLeNet model when images in the validation set are rotated by $0^\circ$, $90^\circ$, $180^\circ$, and $270^\circ$, respectively. In Figure \ref{fig:fig08}, we present the overall accuracy as a function of training steps for the GoogLeNet model applied to the images with and without APCT. The cyan and magenta lines represent the median accuracy rate in each step bin, while the shaded areas enclose the corresponding 16th and 84th percentiles at each step bin. Different panels show the results obtained when the validation set is rotated at different angles.

From this figure, it is evident that regardless of the angle at which the validation set is rotated, preprocessing leads to an overall accuracy approaching 100\%. Moreover, after reaching its peak, the accuracy of the validation set remains nearly constant as the number of training steps increases. In contrast, without preprocessing, the results are comparable to those with preprocessing when the validation set images are not rotated. However, when the validation set is rotated, the overall accuracy decreases by approximately 2.0\% to 6.0\%. Additionally, after reaching its peak, the accuracy of the validation set still exhibits considerable fluctuations as the number of training steps increases. This underscores the effectiveness of our preprocessing and highlights the poor rotational robustness of the traditional CNN framework.

\section{Prepare for CSST} \label{sec:CSST}

The CSST is a 2-meter aperture space telescope positioned in the same orbit as the China Manned Space Station. The CSST's survey camera is outfitted with 18 multi-band imaging detectors, encompassing 7 wavelength bands (NUV, u, g, r, i, z, and y). Each detector offers a field of view spanning \(11 \times 11\) arcmin\(^2\), with a scale of \(9k \times 9k\) pixels, and an average pixel size of $0\farcs74$ \citep{TB-2021-0016}. Scheduled to commence observations in 2026, the CSST is slated for a 10-year operational span. Over its operational period, the CSST aims to image approximately 15,000 square degrees of the sky at a depth of $\rm r=26.0$ mag and 400 square degrees at a depth of $\rm r=27.2$. In anticipation of the CSST's launch, we have evaluated the reliability of our framework using simulated CSST data.

\subsection{Simulation Data}\label{subsec:CSST data}

In this study, we utilize the data from the CSST main sky survey simulation data\footnote{https://csst-tb.bao.ac.cn/code/csst-sims/csst\_msc\_sim.git} to evaluate the performance of GoogLeNet in discriminating between stars and galaxies. The CSST simulation data used in this work is from version C6.2 of the CSST Main Sky Survey Simulation Data. The software used is the CSST Main Sky Survey Simulation Software 2.0.0. For comprehensive details regarding the simulation data, please consult the pertinent website. Briefly, as outlined by \cite{2018MNRAS.480.2178C} and \cite{2023FrASS..1046603F}, the simulations are generated through the following steps: (1) creation of an input catalog containing sufficient physical properties (e.g., magnitude, redshift, source type, source shape, and source size) of all objects utilized in generating CSST observations, and (2) simulation of various physical and instrumental effects during observations, encompassing cosmic rays, sky background, non-linearity, distortion, dark current, flat field, bias, charge diffusion effect, failed image elements/columns, CCD saturation overflow, instrument platform jitter, gain, readout noise, and others. The input galaxy catalog is derived from the  ``JiuTian'' cosmological simulation catalog (e.g., \citealt{2016MNRAS.458..366L, 2018ApJ...853...25W, 2022ApJ...930...66Q}), while the input star catalogs are sourced from Gaia DR3 (e.g., \citealt{REF1, 2021A&A...649A...2L, 2023A&A...674A...1G}) and one simulated catalog obtained with \texttt{Galaxia} \citep{2011ApJ...730....3S}. The final simulation comprises 137 multi-band exposures centered at RA=244.9727 deg, Dec=39.8959 deg, covering an area of approximately 1.53 square degrees. In the subsequent analyses, we focus on the data obtained in the i band, which is analogous to the I band utilized in our previous analysis.

\subsection{Classification Results of CSST Simulation Data}\label{subsec:CSST result}

Similar to Section \ref{sec:data}, we obtain simulated images of stars and galaxies along with their corresponding input catalog. Given the known types of input sources, no additional steps are necessary to ensure a clean sample. We restrict our analysis to objects brighter than 24 mag in the i band when considering that the CSST depth is shallower compared to that of the COSMOS field. For galaxies, we further constrain our sample within the redshift range of $0.2 < z < 1.2$. After excluding images contaminated with cosmic rays, we obtain data for 180,135 galaxies and 34,788 stars. Taking into account the CSST pixel size as $0\farcs074$, we crop these simulated samples into cutouts with a size of $\rm 42\times 42$ pixels. Then we apply preprocessing methods, including noise reduction and APCT, to these images.

\begin{figure}
\centering
\includegraphics[scale=0.3]{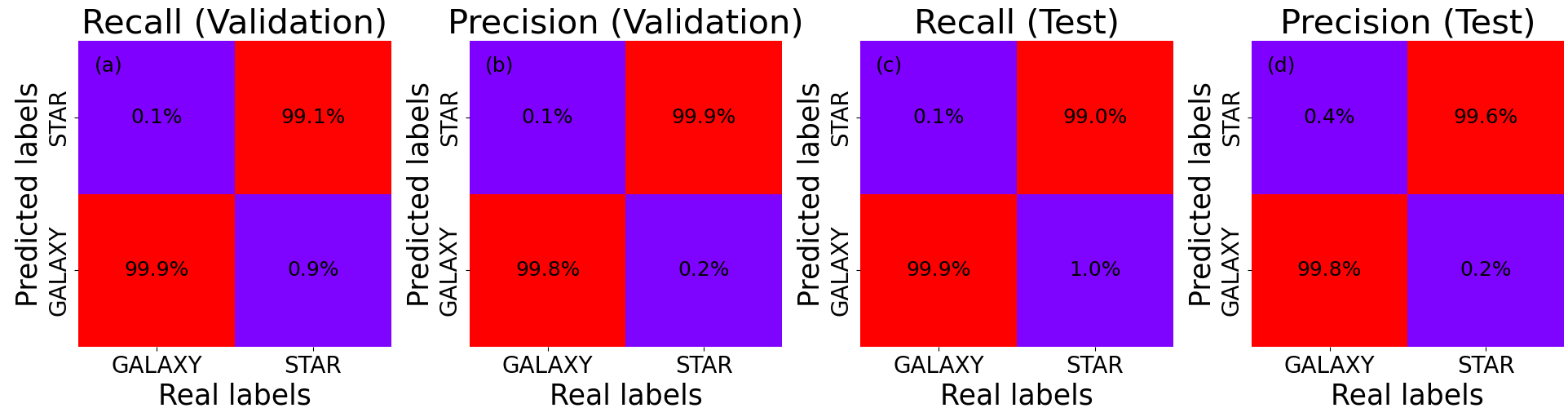}  
\caption{The results obtained from the CSST simulation data are depicted in Panels (a) and (b), illustrating the recall and precision rates derived from the validation set, respectively. Similarly, Panels (c) and (d) present the results obtained from the test set. Across all panels, the overall accuracy rates surpass 99.0\%, indicating the suitability of our framework for future CSST missions.}
\label{fig:fig09}    
\end{figure}

To replicate the workflow after the observation on CSST, we employ data from the initial 45 exposures to train the GoogLeNet network. Subsequently, we utilize the trained network to classify the types of samples acquired from the remaining 92 exposures (referred to as the test set). During the network training phase, we partition the samples into training and validation sets in an 8:2 ratio to avoid overfitting. The detailed breakdown of stars and galaxies in the various datasets is presented in Table \ref{tab:2}. Our classification outcomes are illustrated in Figure \ref{fig:fig09}. Panels (a) and (b) depict the recall and precision rates of the validation set, while panels (c) and (d) exhibit the corresponding results of the test set. Notably, from panels (a) and (b), it is evident that the performance of the GoogLeNet model on the validation set mirrors the outcomes obtained with the COSMOS data, with overall accuracy rates surpassing 99.0\%. Furthermore, when assessing the model's performance on a smaller training dataset and a larger test dataset, the results remain largely consistent. The recall (precision) rate for galaxies on the test set is 99.9\% (99.8\%), while for stars, the corresponding result is 99.0\% (99.6\%). These findings suggest that our network can be trained effectively with a limited dataset and subsequently deployed with high reliability for the forthcoming CSST surveys.


\subsection{Discussion for applications on real CSST
images}\label{subsec:CSST discuss}

In this study, we have demonstrated the effectiveness of our algorithm in classifying galaxies and stars using data from the HST/COSMOS field and CSST simulation data. However, when applying it to future real CSST data, further considerations may be necessary. First, in this study, we only considered relatively bright samples. For the fainter samples, due to their lower signal-to-noise ratio, the classification results may not be as reliable. To address this issue, we need more samples of faint sources with reliable labels. For example, we can obtain reliable labels for more faint samples through simulated data and semi-analytic models. Additionally, CSST will provide 150 square degrees of deep field data, which will also aid in our study of faint sources. In addition, we used only single-band data for our samples. By incorporating multiband images of galaxies, we believe that we can achieve even more reliable results. We are also continuously iterating on our algorithm to make it more effective in extracting characteristic information from faint sources.

In addition, for real images, we do not have an input catalog to provide us labels for training samples. However, fortunately, many excellent instruments (e.g., HST, Euclid, Spitzer, and JWST) have already provided us with many high-quality data. By combining CSST's high-resolution images with multiband photometric data and spectral data from other instruments, we can construct our galaxy and star samples by simultaneously considering the morphology and SED shapes of the targets (just like in Section 2.4 of our study and Section 5.1 of \citealt{2022ApJS..258...11W}).

Moreover, due to CSST's high-efficiency survey capability, we will acquire a substantial amount of data over a short period. This imposes high demands on the efficiency of our data processing, such as how to cut out the stamp images of objects more efficiently given their large number. We have currently significantly improved the efficiency of cutting out these images by using multithreading techniques. However, current method requires us to perform source detection in advance to obtain the coordinates of the targets. In the future, We are considering using some deep learning techniques (e.g., the YOLO algorithm) to automatically detect sources and extract the target galaxies \citep{he2021deep}. This will greatly enhance the efficiency of our entire workflow, making it more suitable for the large amount of data in the future.

\begin{table} 
\centering
\caption{The numbers of stars and galaxies in the training, validation, and test sets.}\label{tab:2} 
\setlength{\tabcolsep}{2mm} 
\begin{tabular}{ccccc} 
\hline\hline\noalign{\smallskip}	
Type & Training set & Validation set & Test set & Total \\ 
\noalign{\smallskip}\hline\noalign{\smallskip}
star & 9276 & 2319 & 23193 & 34788\\
galaxy & 48036 & 12009 & 120090 & 180135\\
\noalign{\smallskip}\hline
\end{tabular} 
\end{table}

\section{Summary} \label{sec:summary}

In this study, we assess the efficacy of the GoogLeNet algorithm in distinguishing between stars and galaxies using data sourced from the COSMOS field, with data preprocessing applied. Through meticulous selection, we procure clean samples of stars and galaxies from the COSMOS2020 catalog. Subsequently, we employ the noise reduction technique with CAE and adopt APCT to transform the images into a polar-coordinate system. Partitioning the samples into training and validation sets in an 8:2 ratio, we observe remarkably high accuracy of the GoogLeNet model on the validation set, surpassing an overall accuracy of 99.0\%. Notably, when the validation set is subjected to rotation, the accuracy attained without preprocessing exhibits a decrease of approximately 2.0\% to 6.0\% compared to the accuracy achieved with preprocessing. This underscores the superiority of our preprocessing approach in mitigating the SML method's poor robustness to image rotation. Moreover, in anticipation of the forthcoming CSST missions, we extend our framework to the CSST simulation data. Remarkably, the GoogLeNet model showcases exceptionally high accuracy, affirming the suitability of our framework for future CSST data analysis.

\begin{acknowledgements}
This work is supported by the Strategic Priority Research Program of Chinese Academy of Sciences (Grant No. XDB 41000000), the National Science Foundation of China (NSFC, Grant No. 12233008, 11973038), the China Manned Space Project (No. CMS-CSST-2021-A07) and the Cyrus Chun Ying Tang Foundations.
Z.S.L. acknowledges the support from Hong Kong Innovation and Technology Fund through the Research Talent Hub program (GSP028).
\end{acknowledgements}

\end{document}